\documentclass[twocolumn,aps,preprintnumbers,amsmath,amssymb]{revtex4}

\usepackage[dvips]{color}
\usepackage{hyperref}

\usepackage{graphics}
\usepackage{graphicx}
\usepackage{dcolumn}
\usepackage{bm}
\usepackage{epsfig} 
\usepackage{psfrag} 

\usepackage[dvips]{color}

\hypersetup{
pdfauthor = {Philippe Curty and Matteo Marsili},
pdftitle = {Phase coexistence in a forecasting game},
pdfsubject = {article},
pdfkeywords = {finance, herding, game theory},
pdfcreator = {LaTeX with hyperref package},
pdfproducer = {dvipdf}
}

\begin{document}


\title{Phase coexistence in a forecasting game}
\author{Philippe Curty and Matteo Marsili}
\affiliation{The Abdus Salam International Center of Theoretical Physics, Trieste, Italy}
\date{\today}

\begin{abstract}
Individual choices are either based on personal
experience or on information provided by peers. The latter case,
causes individuals to conform to the majority in their neighborhood.
Such herding behavior may be very efficient in aggregating disperse
private information, thereby revealing the optimal choice. However if
the majority relies on herding, this mechanism may dramatically fail
to aggregate correctly the information, causing the majority adopting
the wrong choice.  We address these issues in a simple model of
interacting agents who aim at giving a correct forecast of a public
variable, either seeking private information or resorting to herding.
As the fraction of herders increases, the model features a phase
transition beyond which a state where most agents make the correct
forecast coexists with one where most of them are wrong. Simple
strategic considerations suggest that indeed such a system of agents
self-organizes deep in the coexistence region.  There, agents tend to
agree much more among themselves than with what they aim at
forecasting, as found in recent empirical studies.
\end{abstract}
\maketitle

Information affects in many subtle ways socio-economic behavior, giving rise to non-trivial collective phenomena.
For example, a key function of markets is that of aggregating the information scattered among traders into
prices. However, if traders rely on the information conveyed by prices, this same mechanism may lead to self-sustaining speculative bubbles. Likewise, we deduce the worth of a restaurant or the importance
of a research subject from its crowdedness or popularity.
However, popularity can consecrate even totally random choices \cite{Bikhchandani}.

Collective herding phenomena in general pose quite interesting problems in statistical physics. To name a few examples,
anomalous fluctuations in financial markets \cite{cont2000,eguiluz2003,sornette}, opinion dynamics \cite{stauffer,weisbuch2001} and the way in which social changes take place \cite{rfim_bouch}
have been related to percolation and random field Ising models. 
It is natural to expect herding behavior  
when it is convenient for the individuals to follow the herd. 
For example, when the majority is buying in the stock market, 
prices go up, hence buying becomes the right thing to do (at least 
in the short run). If a technology (e.g. fax machine) is widely 
adopted, it becomes more convenient to adopt it.
Herding takes place even in cases where agents' behavior does not 
influence the outcome, if agents try to infer information about the 
optimal choice from the actions of others. Ref. \cite{Bikhchandani} 
discusses the relevance of these considerations for issues ranging 
from the prevalence of crime, marketing, fads and fashions to the 
onset of protests such as that leading to the collapse of the East 
German regime. Ref. \cite{guedj2004} remarks that herding might 
explain why financial forecasters tend to make very similar predictions -- whose 
diversity is much smaller than the prediction's error.

From the theoretical side, the onset of herding and the resulting 
failure of information aggregation has been shown to occur in 
models of {\em information cascades} \cite{Bikhchandani}. The 
prototype example is that of a number of individuals choosing one 
of two restaurants on the basis of some private noisy 
information. If each of them follows the recommendation of 
his/her private signal, the majority will choose the best 
restaurant. However if an individual can observe what others have 
chosen before, he/she can infer their information from their 
choices and take advantage of this. This however leads him/her to 
follow the crowd disregarding private information. As a result, 
choices disclose no further information and there is a sizeable 
probability that all people enter the worse restaurant.  

In this letter, we show that information herding can bring to 
non-trivial collective phenomena even when agents observe a finite 
number of peers and act in no particular order. In particular, a 
population of selfish agents fails to correctly aggregate 
information because herding brings the system into a coexistence 
region, where the vast majority of agents ``agrees'' on the same 
forecast, not necessarily the right one. A statistical mechanics 
approach gives a detailed account of the results in terms of a 
zero temperature spin model with asymmetric interaction. These 
insights extend to the case where agents have to forecast a 
variable in a continuous interval. Again we find a spinodal point 
beyond which forecasts tend to cluster, as observed in Ref. 
\cite{guedj2004}. 

Let us consider a population of agents who have to forecast a 
binary event $E\in\{\pm 1\}$.  Each agent $i=1,\ldots,N$ faces 
the choice of either looking for information or herding. We shall 
denote by $I$ and $H$, respectively, these two strategies, as 
well as the set of agents who follow them. In the former case 
agent $i\in I$ receives some private information $\theta_i\in 
\{\pm 1\}$ about $E$. We assume that $\theta_i$ is drawn 
independently $\forall i\in I$ with $P\{f_i=E\}=p>1/2$, i.e. that 
private signals are informative about $E$. On the basis of this 
signal, agent $i$ makes a forecast $f_i=\theta_i$. In the case of 
strategy $H$, agents receive a private information $\theta_i=\pm 
1$ which is uncorrelated with $E$ (i.e. $P\{\theta_i=\pm 1\}=1/2$ 
i.i.d. for all $i\in H$). Each agent $i\in H$ 
information gathered by a sample group of other agents: He/she 
forms a sample group $G_i$ by picking an odd number $K$ of other 
agents at random, observes their forecasts $f_j$ and sets his/her 
forecast to that of the majority of agents $j\in G_i$. Notice 
first that $j\in G_i$ -- i.e. $i$ observing $j$ -- does not imply 
that $i\in G_j$ -- i.e. that $j$ observes $i$. Secondly, the 
forecast of $i$ may depend on the forecast of other agents who 
are themselves herding. Hence we assume that forecasts are formed 
by the following iterative process, mimiking a sort of 
information exchange: Forecasts are initialized to private 
signals $f_i=\theta_i$ for all $i$. Next, an agent $i\in H$ is 
chosen at random and its forecast is updated to that of the 
majority of $j\in G_i$

\begin{equation}\label{netw}
  f_i\to f_i'={\rm sign}\sum_{j\in G_i} f_j.
\end{equation} 

\noindent 
This process is repeated until it converges and we 
denote simply by $f_j$ the fixed point values of the forecasts. 
It is important to stress that agents receive information on $E$ 
and form their forecast only after they have chosen their strategy.
In other words, agents have access to either type of information but 
not to both. This is natural if both strategies imply a fixed cost: then 
either agents invest in information seeking or in forming a sample group.  

\begin{figure}[ht]
\begin{center}  \setlength\unitlength{1cm}
\begin{picture}(7.5,6)
\put(0,0){\resizebox{7cm}{!}{\includegraphics{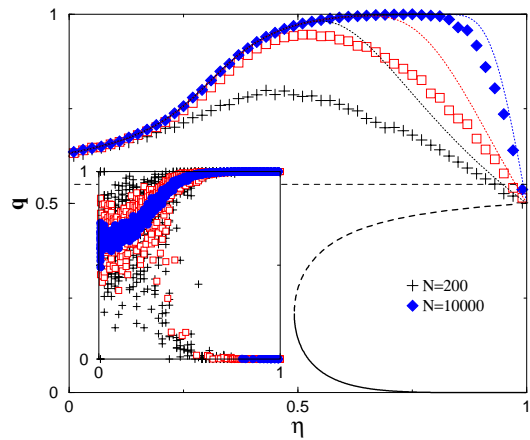}}}
\end{picture}
\end{center}
\vspace{-0.5cm}
\caption{ \label{fig1} The average success $q$ of herding agents is
shown, for simulations (symbols) and for the analytical solution
(dotted lines) as a function of the herding probability $\eta$ for $K=11$,
$p=0.55$ (horizontal line) and $N=200$ ($+$) $10^3$ ($\Box$) and $10^4$ ($\diamond$)
agents. The stable solutions $q_\pm$ are shown as full lines whereas
the unstable one $q_u$ is shown as a dashed line. Inset: individual
realizations of $q$ for the same systems above.}
\end{figure}

Before dealing with the game theoretic case where each agent 
chooses a strategy so as to maximize a payoff, let us focus on 
the case where a fixed fraction $\eta$ of agents follow the $H$ 
strategy and the rest follows the $I$ strategy. By definition, 
the probability of a right forecast of $i\in I$ is $P\{f_i=E|i\in 
I\}=p$, whereas for $i\in H$ we define 
\begin{equation}\label{q_t}
  q\equiv \frac{1}{\eta N}\sum_{i\in H} \delta_{f_i,E} \simeq P\{f_i=E|i\in H\}.
\end{equation}
The inset of Fig. \ref{fig1} shows the behavior of $q$ as a 
function of $\eta$ in typical numerical simulations. The average 
$\langle{q}\rangle$ of $q$ over different realizations is also 
reported in Fig. \ref{fig1}.
When $\eta$ is small, herding is quite efficient and it yields 
more accurate predictions than information seeking ($\langle q\rangle>p$). 
Actually the probability $\langle q\rangle$  that $H$-players end up with 
the correct forecast increases with $\eta$ up to a maximum. This is because 
herders use the information of other herders who have themselves a higher 
performance than private information forecasters. However beyond a certain 
point, outcomes with a value $q<p$ start to appear, coexisting with outcomes 
with $q\approx 1$. Consequently the average $\langle q\rangle$ starts 
decreasing. The low $q$ state becomes more and more probable as $\eta$ 
increases, and for $\eta$ close to one we find $\langle q\rangle <p$.

In order to shed light on the above results, let us notice that the probability of a randomly drawn agent to give the right forecast is
\begin{equation}\label{pi}
P\{f_i=E\}\equiv\pi=(1-\eta)p+\eta q.
\end{equation}
In order to derive an equation for $q$ we observe that a herding agent adopts the point of view of the majority of his $K$ randomly drawn agents, i.e.
\begin{eqnarray}
  q=\Sigma_K(\pi)&\equiv &P\left\{{\rm sign} \sum_{j\in G_i}f_j=E|i\in H\right\}
  \nonumber\\
  &=&\sum_{g=(K+1)/2}^{K}{K \choose g} \pi^g   (1-\pi)^{K-g}\label{theory}
\end{eqnarray}
These are two self consistent equations for $q$. For a given value of
$p$, the solution is unique for $\eta<\eta_c(p,K)$ whereas for
$\eta>\eta_c(p,K)$, as shown in Fig. \ref{fig1},
we find three solutions, which we denote by $q_+>q_u>q_-$. The
critical point $\eta_c$ increases with $p$ and with $K$.

A direct calculation shows that the average number of fixed 
points of Eqs. (\ref{netw}) is dominated by configurations 
$\{f_i\}$ for which $q$ satisfies Eqs. (\ref{pi},\ref{theory}). 
Interestingly, we find that the average number of fixed points 
${\cal N}\simeq (K^K e^{-K}/K!)^{\eta N}$ is the same on all the 
solutions. 


Linear stability shows that the fixed points 
$q_\pm$ are stable whereas the one at $q_u$ is unstable. 
To see this, imagine that at iteration $t$, the fraction of agents 
$i\in H$ with a correct forecast is $q(t)=q^*+\delta q(t)$, 
where $q^*$ is a solution of $q^*=\Sigma_K[\eta q^*+(1-\eta)p]$. 
Then at time $t+1$ we have $\delta q(t+1)\simeq \Sigma_K'\eta \delta q(t)$, and
it is easy to show that $\delta q$ vanishes for $q^*=q_\pm$ whereas it
diverges exponentially for $q^*=q_u$. The unstable solution $q_u$ 
separates the basin of attraction of the fixed points $q_\pm$. 
This allows us to estimate the probability $p_-$ that the system 
converges to the fixed point $q_-$, which is the probability that 
the initial value of $q(0)$ falls below $q_u$. Given that 
variables $\theta_i$ are assigned a random sign for $i\in H$, 
$q^{(0)}$ is well approximated by a gaussian variable of mean 
zero and variance $1/(\eta N)$. Hence 

\begin{equation}\label{pm}
p_-\equiv P\{q(0)<q_u\}\cong \frac{1}{2}{\rm erfc}\left(\sqrt{\eta N/2}(1-2 q_u)\right).
\end{equation}
The expected value of $q$ is then given by
\begin{equation}\label{q}
\langle q\rangle=p_-q_-+(1-p_-)q_+.
\end{equation}
Fig. \ref{fig1} shows that Eq. (\ref{q}) agrees very well with numerical simulations for large $N$. The discrepancy for small $N$ comes from the fact that indeed the dynamics of $q^{(\tau)}$ is
subject to a noise term of order $1/\sqrt{N}$ which causes transitions across $q_u$ in the early stages of the dynamics for small $N$.
It is easy to show that, for $\eta\approx 1$,
\begin{equation}\label{qu}
q_u\simeq \frac{1}{2}-\frac{(p-1/2)k!!}{k!!-(k-1)!!}(1-\eta)+O(1-\eta)^2
\end{equation}
which shows that there is a window of size $1/\sqrt{N}$ close to $\eta=1$ where $p_-$ is sizeable. As a consequence, the fall of $q$ in this region
gets steeper and steeper as $N$ increases.

This consideration is important if we analyze the behavior of selfish
agents following game theory \cite{gt}. We assume for simplicity that
agents aim at reaching a correct forecast, i.e. that their payoff is
the probability that $f_i=E$. As long as $\langle q\rangle>p$ agents will
find it more convenient to switch from the $I$ to $H$ strategy. Hence,
the fraction $\eta$ of herders increases when $\langle q\rangle>p$. The
contrary is true when $\langle q\rangle<p$ and hence we expect that the population
will self-organize to a state $\eta^*$, such that no agent
has incentive to change strategy, i.e. where $\langle q\rangle=p$. Such a state is called a
Nash equilibrium \cite{gt}. Its standard interpretation as the equilibrium of
forward looking rational agents, who correctly anticipate the behavior of others, given the rules of the game,
and respond optimally, requires agents to solve a rather complex
statistical mechanical problem. We will however show below that adaptive agents with limited rationality can
``learn'' to converge to such a Nash equilibrium.

In light of the the results discussed above, the point where $\langle q\rangle 
=p$ -- the Nash equilibrium -- is attained when all but a fraction 
$1-\eta^*\sim N^{-1/2}$ of agents takes the $H$ strategy. 
In addition, because in this region
$q_+\cong 1$ and $q_-\cong 0$, by Eq. (\ref{q})we have $p_-\cong 1-p$. 
This means that
the whole population adopts the wrong forecast with probability $1-p$,
as if it were a single individual forecasting on the basis of private
information.
Such a spectacular event is similar to the outcome of information
cascades \cite{Bikhchandani}, but it takes place in a quite different setting.

Does this scenario changes when we introduce heterogeneity in 
agents' characteristics? Let us first consider the case where 
agent $i$, when using strategy $H$, can observe $K_i$ peers. 
Na\"{i}vely one would expect that agents with larger $K_i$ 
receive more precise information and hence should prefer the $H$ 
strategy. However, because at the Nash equilibrium almost every 
agent is making the same prediction, either right or wrong, a 
larger ``window" $K_i$ does not help. The case where agents have 
different individual forecasting abilities, i.e. when $p_i$ 
depends on $i$, is a bit more complex. It is reasonable to assume 
that ``expert'' agents with $p_i>\langle q\rangle$ will seek 
private information whereas those with $p_i<\langle q\rangle$ 
will herd. Again $q$ is given by Eqs. (\ref{pi},\ref{theory}) with 
\begin{equation}\label{etap} 
\eta=\int_0^{\langle q\rangle}\!dp \ 
\phi(p),~~~~~(1-\eta)p=\int_{\langle q\rangle}^1\!dp \ p \ \phi(p) 
\end{equation} 
where $\phi(p)$ is the distribution of $p_i$. It 
is easy to show that a solution of Eqs. 
(\ref{pi},\ref{theory},\ref{etap}) with $q=\langle q\rangle$, 
i.e. where $\eta$ and $p$ do not fall in the coexistence region 
is not possible. Indeed the only solution of 
$\Sigma_K[q\int_0^q\!dp \phi(p)+\int_q^1\!dp p\phi(p)]=q$ is at 
$q=1$, which implies $\eta=1$. The solution then lies in the 
coexistence region, where Eqs. (\ref{pi},\ref{theory}) have three 
solutions, and it is found computing $\langle q\rangle$ as before 
from Eqs. (\ref{pm},\ref{q}) as a function of $\eta$ and $p$, and 
then using Eq. (\ref{etap}) to compute $\eta$ and $p$ 
self-consistently. Again, the Nash equilibrium lies where 
$p_-\simeq 1-\langle q\rangle$ is finite as $N\to\infty$, 
which, by Eqs. (\ref{pm},\ref{qu}), implies that $\eta^*\to 1$ in 
this limit.

\begin{figure}[ht]
\begin{center}  \setlength\unitlength{1cm}
\begin{picture}(7.5,6)
\put(0,0){\resizebox{7cm}{!}{\includegraphics{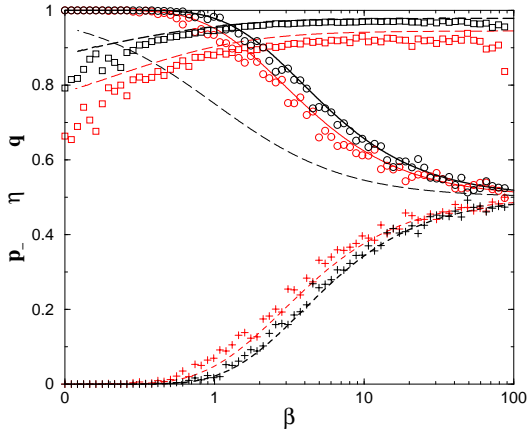}}}
\end{picture}
\end{center}
\vspace{-0.5cm}
\caption{ \label{figbeta} Analytical results (lines) compared to
  numerical simulations (symbols) for systems of $N=100$ and $800$
  agents with heterogeneous forecasting ability $p_i$ drawn from the
  distribution $\phi(p)=\beta 2^\beta(1-p)^{\beta-1}$. The average
  success $q$ (full line and $\circ$), the fraction $\eta$ of herding
  agents (long dashed line and $\Box$) and the probability $p_-$ that
  the majority forecasts the wrong outcome (short dashed line and
  $+$), as a function of $\beta$. For comparison, the thin dashed line
  shows the average success of agents with no herding ($\eta=0$).}
\end{figure}

The results are illustrated in Fig. \ref{figbeta} for the particular case
$\phi(p)=\beta 2^\beta(1-p)^{\beta-1}$, $p\in[1/2,1]$. When 
$\beta$ is large, there is small heterogeneity and $p_i\simeq 1/2$: 
Almost all agents follow the $H$ strategy 
($\eta\approx 1$) and the probability of a wrong forecast $p_-\simeq 
1/2$ is large. As $\beta$ decreases, the number of ``experts'', 
i.e. agents with $p_i>\langle q\rangle$ increases, and correspondingly also the 
performance of the population as a whole improves (i.e. $q$ 
increases and $p_-$ decreases). In this region, asymptotic 
analysis shows that the fraction of ``experts'' $1-\eta\sim 
\sqrt{\log N/N}$.

The analytical results were tested against numerical simulations of
adaptive agents who repeatedly play the game and learn, in the course
of time, about their optimal choice. In order to do this, agents
compute the cumulative payoff for both strategies and adopt the
strategy with the largest score \cite{MG}. As expected, we find that in each run
there is a value $q$ such that all agents with $p_i>q$ play the $I$
strategy whereas those with $p_i<q$ herd. Again some deviations occur
for small $N$ but the agreement improves as $N$ increases. This shows
that the type of equilibria we discuss are ``learnable'' by a
population of not extremely sophisticated agents. It is well known
that the type of reinforcement learning dynamics discussed above has
close analogies with evolutionary dynamics \cite{BorgerSarin}. Hence
the above scenario, might as well describe social norms
which are the result of evolutionary processes.

\noindent
The insights of the discrete model hold also when agents have to
forecast a continuous variable $E$. In order to show this, we
adopt an asymmetric version of the continuous opinion model of Ref. \cite{weisbuch2001}, where a population of $N$ agents submits forecasts $\{f_i\}$ of a continuous
event $E \in [0,1]$. Again, forecasters may either seek private information (strategy $I$) or herd (strategy $H$). All $I$ agents receive a signal $f_i\in [0,1]$ which, with probability $p$ is ``correct'', i.e. is randomly drawn from the interval $[E-\epsilon,E+\epsilon]$, and with probability $1-p$ is uniformly distributed in $[0,1]$.
If instead $i\in H$, we draw at random sample groups $G_i$ of $K$ agents and assign initial random values $f_i^{(0)}\in [0,1]$ to herding agents. Then we iterate the dynamics over agents $j$ of the the group $G_i$
\[
  f_i^{(\tau+1)}=f_i^{(\tau)}+\mu(f_j^{(\tau)}-f_i^{(\tau)})   \ \theta\left(d-|f_j^{(\tau)}-f_i^{(\tau)}|\right)
\]
until $|f_i^{(\tau+1)}-f_i^{(\tau)}|<\epsilon$. We denote simply by $f_i$ the limit value of $f_i^{(\tau)}$ in this process. Note that agent $i$ is influenced by $j\in G_i$ only if their opinion are not too far, i.e. if $|f_j^{(\tau)}-f_i^{(\tau)}|<d$. Forecasts are considered to be correct if $|f_i-E|<\epsilon$.

\begin{figure}
\begin{center}  \setlength\unitlength{1cm}
\begin{picture}(8,6)
\put(0,0){\resizebox{8cm}{!}{\includegraphics{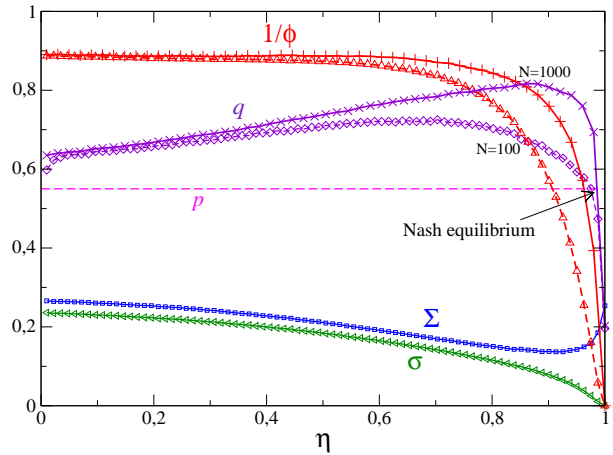}}}
\end{picture}
\end{center}
\vspace{-0.5cm}
\caption{ \label{fig-continuous-forecast}
Continuous forecasting model for $K=11, d =\mu =0.5, \varepsilon=0.1$. The inverse herding parameter $\phi^{-1}$ is only of the order of 0.1 for a strong herding regime near the Nash equilibrium $\eta \approx 0.98$. The dispersion $\sigma$ and the  error $\Sigma$ are only shown for $N=100$. Note that $\eta_{Nash}$ increases with $N$ whereas $\phi_{Nash}$ decreases.}
\end{figure}

As in Ref. \cite{guedj2004}, we introduce the forecast error $\Sigma =  \sqrt{ \langle (\bar{f}-E)^2 \rangle}$ and
the forecast dispersion $\sigma=\sqrt{\langle\overline{(f_i-\bar{f})^2}\rangle}$ where $\bar{\cdot}$ denotes the average over agents whereas the average $\langle \ldots \rangle$ is taken over different realizations of the process. The ratio $\phi = \Sigma / \sigma$ called the empirical herding coefficient, is a measure of herding as explained in Ref.  \cite{guedj2004}.
Fig. \ref{fig-continuous-forecast} shows the results of numerical simulations of the model as a function of the fraction $\eta$ of herders, for a typical choice of the parameters. As in the discrete model, we find that for small values of $\eta$ the probability $q=P\{|f_i-E|<\epsilon|i\in H\}$ of a correct forecast for herders is larger than that of information seeking agents ($p$) and it increases because herding agents aggregate the information of other agents who are also herding. Upon increasing $\eta$ further, $q$ reaches a maximum and then it decreases as the information entering in the system diminishes. In this region, we find coexistence of a state where the vast majority of agents are right with a state where almost all of them are wrong.
The Nash equilibrium, where both strategies are equally successful ($\langle q\rangle =p$), is precisely in this region and the herding coefficient attains values $\phi\simeq 5\div 10$, which are comparable to those found in Ref. \cite{guedj2004} on a survey of earning forecasters of US, EU, UK and JP stocks during the period 1987-2004.
The fact that analysts agree with each other five to
ten times more than with the actual result, was claimed to be related to herding effects in Ref. \cite{guedj2004},
a conclusion fully supported by our results.
Furthermore, as in the discrete model, the Nash equilibrium moves towards $\eta =1$ as $N$ increases, thus making herd behavior more pronounced.

In conclusion, we introduced a simple model capturing the tension
between private information seeking and exploiting information
gathered by others (herding) in a population. When few agents herd,
information aggregation is very efficient.  This makes herding the
choice taken by nearly the whole population, thus setting the system
deep in a ``coexistence'' region where the population as a whole
adopts either the right or the wrong forecast.  This scenario is
rather robust and applies both to a discrete and a continuum model and
it compares well with empirical findings \cite{guedj2004}. The model
and the statistical mechanics analysis can serve as a basis to address
a wide range of related issues.

We are grateful to J.-P. Bouchaud, S. Goyal and F. Vega-Redondo for
useful discussions. We acknowledge financial support from Swiss
National Science Foundation and from EU grant HPRN-CT-2002-00319,
STIPCO and EU-NEST project COMPLEXMARKETS.

\bibliographystyle{h-physrev4}


\begin{thebibliography}{10}

\bibitem{Bikhchandani}
S.~Bikhchandani, D.~Hirshleifer and I.~Welch,
\newblock J. Pol. Econ. {\bf 100} (1992).

\bibitem{cont2000}
R.~Cont and J.~Bouchaud,
\newblock Macroeconomic Dynamics {\bf 4}, 170 (2000).

\bibitem{stauffer}
D.~Stauffer,
\newblock Adv. Complex Syst. {\bf 4} (2001).

\bibitem{weisbuch2001}
G.~Weisbuch and {\it alter},
\newblock Complexity {\bf 7}, 55 (2002).

\bibitem{eguiluz2003}
V.~Egu\'iluz and M.~Zimmermann,
\newblock Phys. Rev. Lett. {\bf 85}, 5659 (2003).

\bibitem{sornette}
W.-X. Zhou~ and D.~Sornette,
\newblock e-print physics {\bf 0503230} (2005).

\bibitem{rfim_bouch}
Q.~Michard and J.-P. Bouchaud,
\newblock cond-mat {\bf 0504079} (2005).

\bibitem{guedj2004}
O.~Guedj and J.-P. Bouchaud,
\newblock cond-mat {\bf 0410079} (2004).

\bibitem{gt}
F.~Vega-Redondo,
\newblock {\em Economics and the theory of games} (Cambridge Univ. Press,
  2004).

\bibitem{MG}
D.~Challet, M.~Marsili and Y.-C. Zhang,
\newblock {\em The Minority Game} (Oxford Univ. Press, 2004).

\bibitem{BorgerSarin}
T.~Borgers and R.~Sarin,
\newblock J. Econ. Th. {\bf 77} (1997).

\end{thebibliography}

\end{document}